\documentstyle[subeqn,epsfig]{elsart}

\newcommand{\rmt}[1]{\tiny\rm #1}

\newcommand{\bm}[1]{\mbox{\boldmath $#1$}}
\newcommand{\bms}[1]{\mbox{\scriptsize\boldmath $#1$}}

\begin{document}
\begin{frontmatter}

%
\title{A formulation for description of $\pi^+(2\pi^-)$ and $\pi^-(2\pi^+)$ channels in Bose-Einstein correlation by Coulomb wave function}
\author[Shinshu,Bohr]{M. Biyajima\thanksref{minoru},}
\author[Toba]{T. Mizoguchi\thanksref{takuya}}
\author[Matsumoto]{and N. Suzuki\thanksref{suzuki}}
\address[Shinshu]{Department of Physics, Faculty of Science, Shinshu 
University, Matsumoto 390-8621, Japan}
\address[Bohr]{Niels Bohr Institute, University of Copenhagen, Denmark}
\address[Toba]{Toba National College of Maritime Technology, Toba 517-8501, 
Japan}
\address[Matsumoto]{Matsumoto University, Matsumoto 390-1295, Japan}
 
\thanks[minoru]{E-mail: biyajima@azusa.shinshu-u.ac.jp}
\thanks[takuya]{E-mail: mizoguti@toba-cmt.ac.jp}
\thanks[suzuki]{E-mail: suzuki@matsu.ac.jp}

\begin{abstract}
%
%
In order to analyze 
data on charged pions correlation channels, $\pi^+(2\pi^-)$ and
$\pi^-(2\pi^+)$, we propose new interferometry approach using 
the Coulomb wave function. 
We show that to describe adequately data
we have to introduce 
 new parameter
describing the contribution of $\pi^-(\bm
k_1)\pi^+(\bm k_2) \rightarrow \pi^-(\bm k_2)\pi^+(\bm k_1)$ process.
Using 
this new formula we analyze data
on $\pi^+(2\pi^-)$ and $\pi^-(2\pi^+)$ channels at $\sqrt s = 91$ GeV by
DELPHI Collaboration, and estimate
the magnitude of this new parameter 
as well as the degree of coherence. 
\end{abstract}
\end{frontmatter}

\section{Introduction}
%
%
In 1995 DELPHI Collaboration reported data of the 3rd order Bose-Einstein
Correlations (BEC) and concluded that there is a genuine the 3rd order
BEC in $3\pi^-$ channel and the effect of the 2nd order BEC in
$\pi^+(2\pi^-)$ channel~\cite{Abreu:1995sq,Biyajima:1990ku}. The
method used in their analyses is the formulation in terms of the
plane wave amplitude.  

From theoretical point of view, at almost the same time, formulations
for the 2nd order BEC by means of the Coulomb wave function had been
investigated in Refs.~\cite{Biyajima:1995ig,Osada:1996cy} 
\footnote{
%
%
For recent presentation of various aspects of BEC in high energy
physics one should consult a recent review \cite{Alexander:2003ug}.
}. Moreover, formulations for the 3rd and 4th order BEC by means of the 
Coulomb wave function have been recently proposed
 in Refs.~\cite{Alt:1998nr,Mizoguchi:2000km}. In particular, the 
formulation of Ref.~\cite{Mizoguchi:2000km} contains the degree of
coherence ($\lambda^{1/2}$) for the magnitude of the BE exchange
term. The advantage 
of this approach
is that it can be directly applicable
to analyses of the data on $3\pi^-$ channel with the CERN-MINUIT
program.\footnote{ 
%
%
In application of formulas in Ref.~\cite{Alt:1998nr} we have to
assume the parameter of the interaction region ($R$) a priori.}  

In the present study we investigate whether
this approach using formula derived by means of the Coulomb wave
function can be extended 
to an unlike charged 
combination channel,  $\pi^+(2\pi^-)$. We should stress that:
\begin{itemize}
  \item[1)] Through our approach, the plane wave amplitude is used as
a basic calculation. At the second step the Coulomb wave function is
utilized. 
  \item[2)] The interferometry effect for $\pi^+(2\pi^-)$,
i.e., the squared amplitude by the Coulomb wave function, contains the
following neutral current:
\begin{eqnarray}
  \pi^-(\bm k_1)\pi^+(\bm k_2) \longrightarrow \pi^-(\bm k_2)\pi^+(\bm k_1)\ .
  \label{eq1}
\end{eqnarray}
If the magnitude of this contribution is zero, we cannot explain the
data on $\pi^+(2\pi^-)$ channel by DELPHI Collaboration. Their data
ask for the finite
magnitude which is expressed by a new parameter $\beta$ in this paper.

\end{itemize}
This paper is organized in the following way:
In \S 2, a simple model for $\pi^+(2\pi^-)$
channel is investigated. In \S 3, a model including full amplitudes
($3!=6$) is presented. A new formula for $\pi^+(2\pi^-)$ channel is
also derived here.
In \S 4, analyses of the data on $3\pi^-$ channel are
performed. The estimated parameters are compared with those in
$\pi^+(2\pi^-)$ channel. In the final section, concluding remarks are
given. 

\section{A simple model}
\begin{figure}
%
%
  \centering
  \epsfig{file=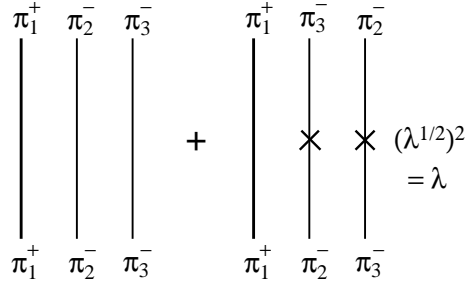,scale=0.6}
  \caption{A thick line denotes the positive pion. Thin lines do negative pion. The cross mark (${\large \times}$) means the exchange effect due to the Bose-Einstein statistics, whose magnitude is expressed by the effective degree of coherent ($\lambda^{1/2}$). }
  \label{fig1}
\end{figure}

%
%
Authors
 of Ref.~\cite{Abreu:1995sq} stressed that data on $\pi^+(2\pi^-)$
channel are described 
by the formulas of the 2nd order BEC. We shall study whether this statement
is correct or not using the interferometry approach formulated in terms of
the Coulomb wave function.
To analyze data on BEC effect in $e^+e^- \to \pi^+(2\pi^-) + X$
channel, first of all we consider the simplest diagram shown in
Fig.~\ref{fig1}.  

The plane wave amplitudes ($PWA$) for $\pi^+(2\pi^-)$ channel are given as 
\begin{subeqnarray}
  \label{eq2}
  PWA(1;\,+--) &=& \frac 1{\sqrt 2}
  e^{ i (\bms k_1^{(+)} \cdot \bms x_1
       + \bms k_2 \cdot \bms x_2
       + \bms k_3 \cdot \bms x_3)}\ ,\\
  PWEA(1;\,+--) &=& \frac 1{\sqrt 2}
  e^{ i (\bms k_1^{(+)} \cdot \bms x_1
       + \bms k_2 \cdot \bms x_3
       + \bms k_3 \cdot \bms x_2)}\ ,
\end{subeqnarray}
where $(+)$ stands for the positive pion and $PWEA$ 
represents the plane wave exchange amplitude due to the BE statistics.

When the Coulomb wave functions is taken into account because of charged pions ($\pi^+(2\pi^-)$), the above $PWA$ and $PWEA$ should be replaced by $A(1;\,+--)$ and $EA(1;\,+--)$, 
\begin{subeqnarray}
\label{eq3}
  A(1;\,+--) &=& 
  \psi_{\bms k_1\bms k_2}^{C+-}(\bm x_1,\ \bm x_2)
  \psi_{\bms k_2\bms k_3}^{C--}(\bm x_2,\ \bm x_3)
  \psi_{\bms k_3\bms k_1}^{C+-}(\bm x_3,\ \bm x_1)\ ,\\
  EA(1;\,+--) &=& 
  \psi_{\bms k_1\bms k_2}^{C+-}(\bm x_1,\ \bm x_3) 
  \psi_{\bms k_2\bms k_3}^{C--}(\bm x_3,\ \bm x_2) 
  \psi_{\bms k_3\bms k_1}^{C+-}(\bm x_2,\ \bm x_1)\ ,
\end{subeqnarray}
where $\psi_{\bms k_1\bms k_2}^{C}$ is defined as
\begin{eqnarray}
  \psi_{\bms k_i \bms k_j}^C(\bm x_i,\ \bm x_j) = \Gamma(1 + i\eta_{ij})
  e^{\pi \eta_{ij}/2} e^{ i\bms k_{ij} \cdot \bms r_{ij} }
  F[- i \eta_{ij},\,1;\,i ( k_{ij} r_{ij} - \bm k_{ij} \cdot \bm r_{ij} )] \qquad
  \label{eq4}
\end{eqnarray}
with $\bm r_{ij} = \bm x_i - \bm x_j$, $\bm k_{ij} = (\bm k_i - \bm k_j)/2$, $r_{ij} = |\bm r_{ij}|$, $k_{ij} = |\bm k_{ij}|$ and $\eta_{ij} = \pm m\alpha/k_{ij}$ (($+$) and ($-$) are the like-charge combination and unlike-charge one, respectively). $F[a,\ b;\ x]$ and $\Gamma(x)$ are the confluent hypergeometric function and the gamma function, respectively. The parts of the plane wave of Eq.~(\ref{eq3}) are given as~\cite{Schiff:1955aa,Sasakawa:1991aa}
\begin{subeqnarray}
\label{eq5}
  A(1;\,+--) && \stackrel{\rmt{PWA}}{\longrightarrow} e^{(3/2)i [\bms k_1^{(+)}\cdot\bms x_1 + \bms k_2\cdot\bms x_2 + \bms k_3\cdot\bms x_3]}\ ,\\
  EA(1;\,+--) && \stackrel{\rmt{PWA}}{\longrightarrow} e^{(3/2)i [\bms k_1^{(+)}\cdot\bms x_1 + \bms k_3\cdot\bms x_2 + \bms k_2\cdot\bms x_3]}\ ,
\end{subeqnarray}
where the factor ($3/2$) is attributed to property of the Coulomb wave function and 3-body problem.

The interferometry effect for the ($\pi^+(2\pi^-)$) channel is calculated as
\begin{eqnarray}
  \frac{N^{(\pi^+2\pi^-)}}{N^{BG}} = (1+\gamma Q_3) \prod_{i=1}^3 \int \rho(\bm x_i) d^3 \bm x_i \left[F_1^{(--+)} + \lambda F_2^{(--+)}\right]\ ,
  \label{eq6}
\end{eqnarray}
where $\rho(\bm x_i)$ stand for the source functions of particle $i$. We use the Gaussian distribution of the radius $R$, $\rho(\bm x_i)=\frac{1}{(2\pi R^2)^{3/2}} \exp\left[-\frac{{\bm x}^2}{2R^2}\right]$, and
\begin{subeqnarray}
\label{eq7}
&&Q_3 = \sqrt{(\bm k_1 - \bm k_2)^2 + (\bm k_2 - \bm k_3)^2 + (\bm k_3 - \bm k_1)^2}\ ,\\
&&F_1^{(+--)} = \frac 12 [|A(1;\,+--)|^2 + |EA(1;\,+--)|^2]\ ,\\
&&F_2^{(+--)} = {\rm Re}[A(1;\,+--)EA^*(1;\,+--)]\ .
\end{subeqnarray}
The parameter $\lambda^{1/2}$ is introduced  in order to
estimate the strength of the BE effect, where the magnitude of the cross
mark (${\large \times}$) is expressed by $\lambda^{1/2}$. Of course
$\lambda$ should be less than one (1), because it can be interpreted
as the degree of coherence in quantum optics [See also
Ref.~\cite{Mizoguchi:2000km}]. 

\begin{figure}
%
%
  \centering
  \epsfig{file=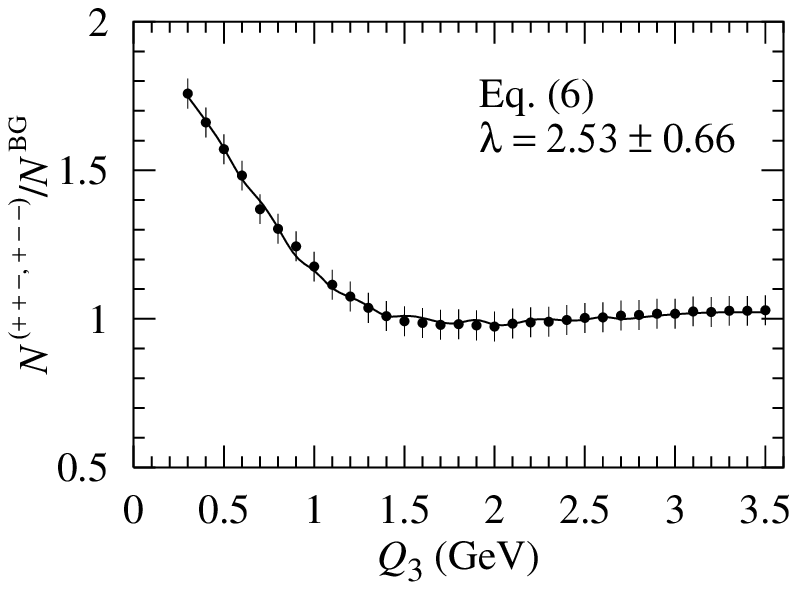,scale=1.2}
  \caption{Analyses of data on $\pi^+(2\pi^-)$ and $\pi^-(2\pi^+)$ channels by means of Eq.~(\ref{eq6}). $\chi^2/n.d.f. = 3.2/29$.}
  \label{fig2}
\end{figure}

Our result is given in Fig. \ref{fig2} and Table \ref{table1}. 
As seen there, the magnitude $\lambda$ is larger than 1. This 
suggests that we have to consider also
additional diagrams (contributions) to Fig.~\ref{fig1}. 
 Therefore we should
seek other possible schemes for the description of
$\pi^+(2\pi^-)$ channel.  

\begin{table}
%
%
  \centering
  \caption{Analyses of $\pi^+(2\pi^-)$ and $\pi^-(2\pi^+)$ BEC by
DELPHI Collaboration. The systematic errors for all points are
assumed to be $\pm 0.05$.
 Small normalizations ($C$) are attributed to
the long range effect $(1+\gamma Q_3)$.}\smallskip 
  \label{table1}
  \begin{tabular}{ccccccc}
  \hline
  formulas
  & $\beta$ (fixed)& $C$ & $R$ [fm]
  & $\lambda$ & $\gamma$ & $\chi^2/N_{dof}$\\
  \hline
  Eq. (\ref{eq6}) 
  & --- & 0.51$\pm$0.11 & 0.13$\pm$0.01
  & 2.53$\pm$0.66 & 0.20$\pm$0.09 & 3.2/29\\
  Eq. (\ref{eq10})
  & 0.8 & 0.79$\pm$0.09 & 0.16$\pm$0.02
  & -0.12$\pm$0.09 & 0.09$\pm$0.05 & 1.3/29\\
  & 0.5 & 0.74$\pm$0.09 & 0.14$\pm$0.01
  & 0.23$\pm$0.12 & 0.11$\pm$0.05 & 1.2/29\\
  & 0.4 & 0.72$\pm$0.09 & 0.13$\pm$0.01
  & 0.39$\pm$0.14 & 0.11$\pm$0.05 & 1.1/29\\
  & 0.3 & 0.70$\pm$0.09 & 0.13$\pm$0.02
  & 0.62$\pm$0.17 & 0.12$\pm$0.05 & 1.1/29\\
  & 0.2 & 0.66$\pm$0.09 & 0.13$\pm$0.01
  & 0.96$\pm$0.23 & 0.13$\pm$0.05 & 1.3/29\\
  & 0.1 & 0.60$\pm$0.10 & 0.13$\pm$0.01
  & 1.51$\pm$0.36 & 0.16$\pm$0.07 & 1.8/29\\
  \hline
  \end{tabular}
\end{table}

\section{A model including six amplitudes}
%
%
As explained in the previous section, we should consider more complex diagrams than Fig.~\ref{fig1}. 
\begin{figure}
%
%
  \centering
  \epsfig{file=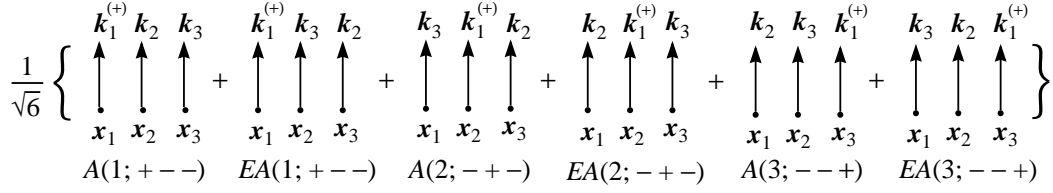,scale=0.57}
  \caption{Six diagrams for three charged pions, because of $3! = 6$.
The BE statistics is taken into account between $\pi_2^{(-)}(\bm
k_2)$ and $\pi_3^{(-)}(\bm k_3)$ pions.} 
  \label{fig3}
\end{figure}
In this case the following $PWA$ and $PWEA$ in addition to
$PWA(1;+--)$ and $PWEA(1;+--)$ are necessary, because of the increased number of diagrams ($3!=6$), 
\begin{subeqnarray}
  \label{eq8}
  PWA(2;\,-+-) &=& 
  e^{ i (\bms k_1^{(+)} \cdot \bms x_2
       + \bms k_2 \cdot \bms x_3
       + \bms k_3 \cdot \bms x_1)}\ ,\\
  PWEA(2;\,-+-) &=& 
  e^{ i (\bms k_1^{(+)} \cdot \bms x_2
       + \bms k_2 \cdot \bms x_1
       + \bms k_3 \cdot \bms x_3)}\ ,\\
  PWA(3;\,--+) &=& 
  e^{ i (\bms k_1^{(+)} \cdot \bms x_3
       + \bms k_2 \cdot \bms x_1
       + \bms k_3 \cdot \bms x_2)}\ ,\\
  PWEA(3;\,--+) &=& 
  e^{ i (\bms k_1^{(+)} \cdot \bms x_3
       + \bms k_2 \cdot \bms x_2
       + \bms k_3 \cdot \bms x_1)}\ .
\end{subeqnarray}
These additional amplitudes described by means of the Coulomb wave function are given as

\begin{subeqnarray}
\label{eq9}
  A(2;\,-+-) &=& 
  \psi_{\bms k_1\bms k_2}^{C+-}(\bm x_2,\ \bm x_3)
  \psi_{\bms k_2\bms k_3}^{C--}(\bm x_3,\ \bm x_1)
  \psi_{\bms k_3\bms k_1}^{C+-}(\bm x_1,\ \bm x_2)\ ,\\
  EA(2;\,-+-) &=& 
  \psi_{\bms k_1\bms k_2}^{C+-}(\bm x_2,\ \bm x_1)
  \psi_{\bms k_2\bms k_3}^{C--}(\bm x_1,\ \bm x_3)
  \psi_{\bms k_3\bms k_1}^{C+-}(\bm x_3,\ \bm x_2)\ ,\\
  A(3;\,--+) &=& 
  \psi_{\bms k_1\bms k_2}^{C+-}(\bm x_3,\ \bm x_1)
  \psi_{\bms k_2\bms k_3}^{C--}(\bm x_1,\ \bm x_2)
  \psi_{\bms k_3\bms k_1}^{C+-}(\bm x_2,\ \bm x_3)\ ,\\
  EA(3;\,--+) &=& 
  \psi_{\bms k_1\bms k_2}^{C+-}(\bm x_3,\ \bm x_2) 
  \psi_{\bms k_2\bms k_3}^{C--}(\bm x_2,\ \bm x_1) 
  \psi_{\bms k_3\bms k_1}^{C+-}(\bm x_1,\ \bm x_3)\ .
\end{subeqnarray}
\begin{figure}
%
%
  \centering
  \epsfig{file=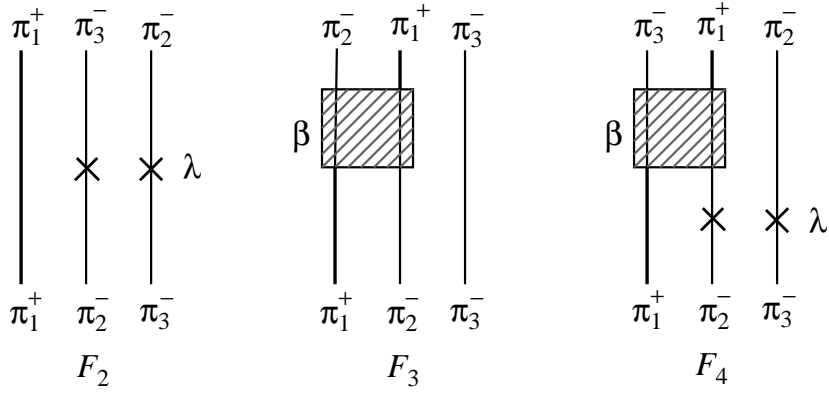,scale=0.7}
  \caption{Diagram-like interpretation of the interferometry 
effect in
Eq.~(\ref{eq10}).} 
  \label{fig4}
\end{figure}
The interferometry effect for $\pi^+(2\pi^-)$ channel described
by Eqs.~(\ref{eq3}) and (\ref{eq9}) is given as  
\begin{eqnarray}
  \frac{N^{(\pi^+2\pi^-)}}{N^{BG}} &=& (1+\gamma Q_3) \prod_{i=1}^3 \int \rho(\bm x_i) d^3 \bm x_i\nonumber\\
  && \times\left[F_1^{(+--)} + \lambda F_2^{(+--)} + \beta F_3^{(+--)} + \lambda \beta F_4^{(+--)}\right]\ ,
  \label{eq10}
\end{eqnarray}
where
\begin{subeqnarray}
  \label{eq11}
  F_1^{(+--)} &=& \frac 16 \left\{\sum_{i = 1}^3 A(i;\,c_1c_2c_3)A^*(i;\,c_1c_2c_3) + \sum_{i=1}^3 EA(i;\,c_1c_2c_3)EA^*(i;\,c_1c_2c_3)\right\}\nonumber\\
  && \stackrel{\rmt{PWA}}{\longrightarrow} 1\ ,\\
  F_2^{(+--)} &=& \frac 16 [ \sum_{i = 1}^3 A(i;\,c_1c_2c_3)EA^*(i;\,c_1c_2c_3) + {\rm c.\ c.}]\nonumber\\
  && \stackrel{\rmt{PWA}}{\longrightarrow} ({\rm typical}\ PW)\ e^{(3/2)i [(\bms k_2-\bms k_3) \cdot \bms x_2 + (\bms k_3-\bms k_2) \cdot \bms x_3]}\ ,\\
  F_3^{(+--)} &=& \frac 16 [ \sum_{i = 1}^3\sum_{j \ne i} A(i;\,c_1c_2c_3)EA^*(j;\,c_1c_2c_3) + {\rm c.\ c.}]\nonumber\\
  && \stackrel{\rmt{PWA}}{\longrightarrow} ({\rm typical}\ PW)\ e^{(3/2)i [(\bms k_1-\bms k_3) \cdot \bms x_1 + (\bms k_3-\bms k_1) \cdot \bms x_3]}\ ,\\
  F_4^{(+--)} &=& \frac 16 [ \sum_{i = 1}^3\sum_{j \ne i} A(i;\,c_1c_2c_3)A^*(j;\,c_1c_2c_3)\nonumber\\
&& \quad + \sum_{i = 1}^3\sum_{j \ne i} EA(i;\,c_1c_2c_3)EA^*(j;\,c_1c_2c_3) ]\nonumber\\
  && \stackrel{\rmt{PWA}}{\longrightarrow} ({\rm typical}\ PW)\ e^{(3/2)i [(\bms k_1-\bms k_3) \cdot \bms x_1 + (\bms k_3-\bms k_2) \cdot \bms x_2 + (\bms k_2-\bms k_1) \cdot \bms x_3]}\ ,
\end{subeqnarray}
In the above equations
for the sake of simplicity we neglect the suffix $(+)$. 
In actual analyses, however we should fix the charge assignment of $(+)$ in $Q_3$. 
Moreover, we have to introduce a new parameter ($\beta$) to
describe the strength of the shadow parts in $F_3^{(+--)}$ and
$F_4^{(+--)}$, cf., Fig.~\ref{fig4}. A possible interpretation
of the role of the shadow region is a ``$\rho^0$-meson-like
contribution'' 
occurring here in order to satisfy the conservation of the
neutral current, cf., Fig.~\ref{fig5}. 
\begin{figure}
%
%
  \centering
  \epsfig{file=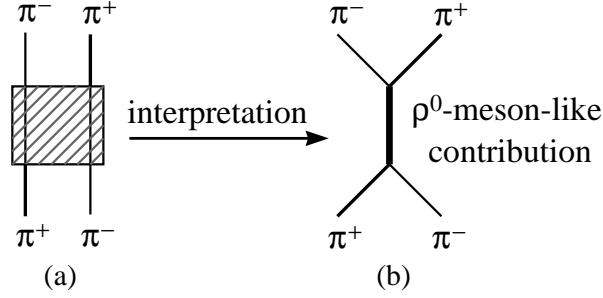,scale=0.7}
  \caption{(a) The exchange part between $\pi^+$ and $\pi^-$
contained in $F_3$. (b) A possibly effective interpretation for the
shadow region in (a) is a ``$\rho^0$-meson-like contribution''.} 
  \label{fig5}
\end{figure}

\begin{figure}
%
%
  \centering
  \epsfig{file=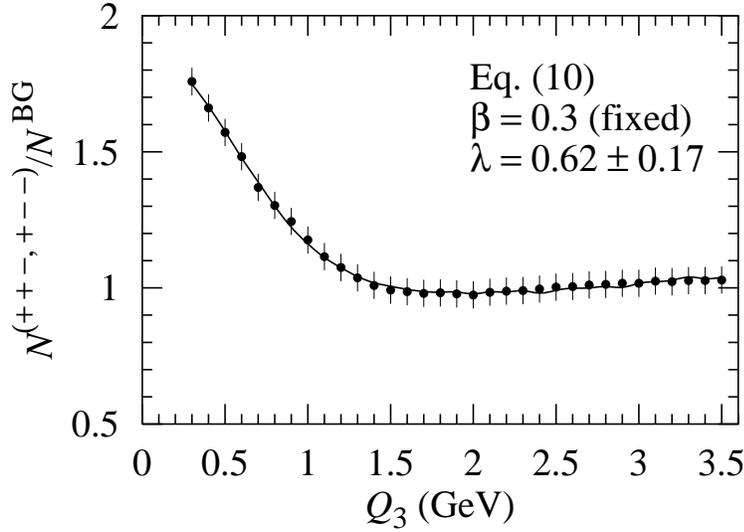,scale=1.2}
  \caption{Analyses of the data on $\pi^+2\pi^-$ and $\pi^-2\pi^+$ channels by means of Eq.~(\ref{eq10}). $\chi^2/n.d.f. = 1.1/29$.}
  \label{fig6}
\end{figure}

The results obtained by our new formula, i.e.,
Eq.~(\ref{eq10}), are given in Fig. \ref{fig6} and Table
\ref{table1}.  

As seen in Table~\ref{table1}, 
$\lambda$ becomes negative for large $\beta$ and exceeds unity for
small $\beta$. The possibly sets of parameters $(\beta,\,\lambda)$
leading to reasonable results are $(\beta,\,\lambda)\sim
(0.5,\,0.2)$, $(0.4,\,0.4)$, $(0.3,\,0.6)$ and $(0.2,\,0.96)$. To
estimate the strength of $\lambda$ we shall now analyze BEC data on
$3\pi^- + 3\pi^+$. 

\section{Analyses of $3\pi^- + 3\pi^+$ BEC by Coulomb wave function}
\begin{figure}
%
%
  \centering
  \epsfig{file=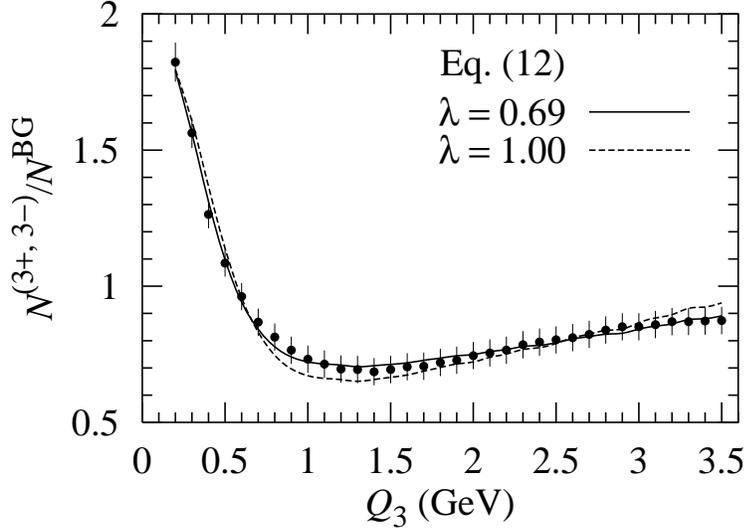,scale=1.2}
  \caption{Analyses of data on $3\pi^+$ and $3\pi^-$ channels by
means of Eq.~(\ref{eq12}).} 
  \label{fig7}
\end{figure}

%
%
To analyze data on 3$\pi^-$ BEC we can use the following formula
presented in Ref.~\cite{Mizoguchi:2000km} 
\begin{eqnarray}
  \frac{N^{(3\pi^-)}}{N^{BG}} = (1+\gamma Q_3) \prod_{i=1}^3 \int \rho(\bm x_i) d^3 \bm x_i \left[F_1^{(3\pi^-)} + \lambda F_2^{(3\pi^-)} + \lambda^{3/2} F_3^{(3\pi^-)}\right]\,,\qquad
  \label{eq12}
\end{eqnarray}
where
\begin{subeqnarray}
  \label{eq13}
  F_1^{(3\pi^-)} &=& \frac 16 \sum_{i=1}^6 A(i)A^*(i) \ ,\\
  F_2^{(3\pi^-)} &=& \frac 16 [ A(1)A^*(2) + A(1)A^*(3) + A(1)A^*(4) + A(2)A^*(5)\nonumber\\
  && \quad  + A(2)A^*(6) + A(3)A^*(5) + A(3)A^*(6) + A(4)A^*(5)\nonumber\\
  && \quad  + A(4)A^*(6) + {\rm c.\ c.}]\ ,\\
  F_3^{(3\pi^-)} &=& \frac 16 [ A(1)A^*(5) + A(1)A^*(6) + A(2)A^*(3) + A(2)A^*(4)\nonumber\\
  && \quad  + A(3)A^*(4) + A(5)A^*(6) + {\rm c.\ c.}]\ ,
\end{subeqnarray}
and
\begin{subeqnarray}
  \label{eq14}
  A(1) &=& 
  \psi_{\bms k_1\bms k_2}^C(\bm x_1,\ \bm x_2)
  \psi_{\bms k_2\bms k_3}^C(\bm x_2,\ \bm x_3)
  \psi_{\bms k_3\bms k_1}^C(\bm x_3,\ \bm x_1)\nonumber\\
  && \stackrel{\rmt{PWA}}{\longrightarrow}
  e^{ i \bms k_{12} \cdot \bms r_{12}}
  e^{ i \bms k_{23} \cdot \bms r_{23}}
  e^{ i \bms k_{31} \cdot \bms r_{31}}
  = e^{ (3/2)i (\bms k_1 \cdot \bms x_1
          + \bms k_2 \cdot \bms x_2
          + \bms k_3 \cdot \bms x_3)}\ ,\\
A(2) &=& 
  \psi_{\bms k_1\bms k_2}^C(\bm x_1,\ \bm x_3) 
  \psi_{\bms k_2\bms k_3}^C(\bm x_3,\ \bm x_2) 
  \psi_{\bms k_3\bms k_1}^C(\bm x_2,\ \bm x_1)\nonumber\\
  && \stackrel{\rmt{PWA}}{\longrightarrow} 
  e^{ i \bms k_{12} \cdot \bms r_{13}} 
  e^{ i \bms k_{23} \cdot \bms r_{32}} 
  e^{ i \bms k_{31} \cdot \bms r_{21}}
  = e^{ (3/2)i (\bms k_1 \cdot \bms x_1
          + \bms k_2 \cdot \bms x_3
          + \bms k_3 \cdot \bms x_2)}\ ,\\
  A(3) &=& 
  \psi_{\bms k_1\bms k_2}^C(\bm x_2,\ \bm x_1)
  \psi_{\bms k_2\bms k_3}^C(\bm x_1,\ \bm x_3)
  \psi_{\bms k_3\bms k_1}^C(\bm x_3,\ \bm x_2)\nonumber\\
  && \stackrel{\rmt{PWA}}{\longrightarrow}
  e^{ i \bms k_{12} \cdot \bms r_{21}}
  e^{ i \bms k_{23} \cdot \bms r_{13}}
  e^{ i \bms k_{31} \cdot \bms r_{32}}
  = e^{ (3/2)i (\bms k_1 \cdot \bms x_2
          + \bms k_2 \cdot \bms x_1
          + \bms k_3 \cdot \bms x_3)}\ ,\\
  A(4) &=& 
  \psi_{\bms k_1\bms k_2}^C(\bm x_2,\ \bm x_3) 
  \psi_{\bms k_2\bms k_3}^C(\bm x_3,\ \bm x_1) 
  \psi_{\bms k_3\bms k_1}^C(\bm x_1,\ \bm x_2)\nonumber\\
  && \stackrel{\rmt{PWA}}{\longrightarrow}
  e^{ i \bms k_{12} \cdot \bms r_{23}}
  e^{ i \bms k_{23} \cdot \bms r_{31}}
  e^{ i \bms k_{31} \cdot \bms r_{12}}
  = e^{ (3/2)i (\bms k_1 \cdot \bms x_2
          + \bms k_2 \cdot \bms x_3
          + \bms k_3 \cdot \bms x_1)}\ ,\\
  A(5) &=& 
  \psi_{\bms k_1\bms k_2}^C(\bm x_3,\ \bm x_1)
  \psi_{\bms k_2\bms k_3}^C(\bm x_1,\ \bm x_2)
  \psi_{\bms k_3\bms k_1}^C(\bm x_2,\ \bm x_3)\nonumber\\
  && \stackrel{\rmt{PWA}}{\longrightarrow}
  e^{ i \bms k_{12} \cdot \bms r_{31}}
  e^{ i \bms k_{23} \cdot \bms r_{12}}
  e^{ i \bms k_{31} \cdot \bms r_{23}}
  = e^{ (3/2)i (\bms k_1 \cdot \bms x_3
          + \bms k_2 \cdot \bms x_1
          + \bms k_3 \cdot \bms x_2)}\ ,\\
  A(6) &=& 
  \psi_{\bms k_1\bms k_2}^C(\bm x_3,\ \bm x_2) 
  \psi_{\bms k_2\bms k_3}^C(\bm x_2,\ \bm x_1) 
  \psi_{\bms k_3\bms k_1}^C(\bm x_1,\ \bm x_3)\nonumber\\
  && \stackrel{\rmt{PWA}}{\longrightarrow} 
  e^{ i \bms k_{12} \cdot \bms r_{32}} 
  e^{ i \bms k_{23} \cdot \bms r_{21}} 
  e^{ i \bms k_{31} \cdot \bms r_{13}}
  = e^{ (3/2)i (\bms k_1 \cdot \bms x_3
          + \bms k_2 \cdot \bms x_2
          + \bms k_3 \cdot \bms x_1)}\ .
\end{subeqnarray}
The result obtained by this formula is given in Fig. \ref{fig7}
and Table \ref{table2}. Notice that 
fixed $\lambda =1$ results in 
large $\chi^2$. This means that 
an additional parameter is necessary here, i.e., that we
should allow the degree of coherence ($\lambda$) to vary as well.
As seen in Table~\ref{table2} we have found that there is a common
region $(\beta,\,\lambda) \sim (0.28,\,0.7)$ for $\pi^+(2\pi^-)$
channel and $\lambda \sim 0.7$ for $3\pi^- + 3\pi^+$ channel. 

\begin{table}
%
%
  \centering
  \caption{Analyses of $3\pi^-$ and $3\pi^+$ BEC by DELPHI
Collaboration. The systematic errors for all points are assumed 
to be 
$\pm 0.05$. Small normalizations ($C$) are attributed to the long
range effect $(1+\gamma Q_3)$.}\smallskip 
  \label{table2}
  \begin{tabular}{ccccccc}
  \hline
  formulas
  & $\beta$ & $C$ & $R$ [fm]
  & $\lambda$ & $\gamma$ & $\chi^2/N_{dof}$\\
  \hline
  Eq. (\ref{eq12}) 
  & --- & 0.33$\pm$0.02 & 0.22$\pm$0.01 
  & 1.0 (fixed) & 0.51$\pm$0.05 & 20.6/31\\
  Eq. (\ref{eq12}) 
  & --- & 0.49$\pm$0.04 & 0.24$\pm$0.01 
  & 0.69$\pm$0.06 & 0.22$\pm$0.05 & 4.0/30\\
  \hline
  \multicolumn{7}{c}{$\pi^+(2\pi^-)$ and $\pi^-(2\pi^+)$ BEC}\\
  Eq. (\ref{eq10}) 
  & 0.28$\pm$0.06 & 0.69$\pm$0.07 & 0.13$\pm$0.01 
  & 0.7 (fixed) & 0.12$\pm$0.04 & 1.2/29\\
  \hline
  \end{tabular}
\end{table}

\section{Concluding remarks}
%
%
We have obtained the new formula for $\pi^+(2\pi^-)$ channel in the
unlike-3rd order BEC, introducing the degree of coherence ($\lambda$)
and the effective magnitude of neutral current ($\beta$). We have
analyzed BEC data on $\pi^+(2\pi^-)$ and $\pi^-(2\pi^+)$ channels in
$e^+e^-$ collision at $\sqrt s = 91$ GeV, using the new formula. 
(Notice that to compare the obtained here values of $R$ with
those obtained by using the plane wave approach, we have to multiply
it by the factor $3/2$, i.e., $R^{(3\pi^-)} \to 0.36$ fm.)

As seen from Table~\ref{table2}, the following choice of parameters
is possible which leads to good fit to data,
$$
(\beta,\,\lambda) \sim (0.28,\,0.7)\ ,
$$ 
provided 
that the degree of coherence ($\lambda$) is almost the same
in both channels. This finite $\beta$ suggests that there is the
genuine 3rd order contribution even in $\pi^+(2\pi^-)$
channel.\footnote{ 
%
%
Hopefully, in future some additional different data on $\pi^+(2\pi^-)$
channel would be available, in which case the usefulness of
Eq.~(\ref{eq10}) could be checked again.}
 To confirm the choice of this set of parameters we need other
data at $\sqrt s = 91$ GeV as well as at different energies.
Moreover, our formula i.e., Eq.~(\ref{eq10}) can be also 
applied to the same kind data from heavy-ion collisions. 

\section*{Acknowledgements}
%
%
One of authors (M. B.) would like to thank The Scandinavia-Japan Sasakawa Foundation for financial support, and authors are also thankful for G.~Wilk's reading the manuscript. They are indebted to useful conversations with H.~Boggild, T.~Csorgo, B.~Lorstad and L.~Muresan.

\end{document}